\documentclass[prd,aps,showpacs,superscriptaddress,nofootinbib,amsmath,amssymb]{revtex4}

\usepackage{graphicx}
\usepackage{dcolumn}
\usepackage{bm}
\usepackage{lipsum}
\begin{document}
\title{Nuclear medium effects in Drell-Yan process}
\author{H. Haider}
\author{M. Sajjad Athar}
\email{sajathar@gmail.com}
\author{S. K. Singh}
\affiliation{Department of Physics, Aligarh Muslim University, Aligarh - 202 002, India}
\author{I. \surname{Ruiz Simo}}
\affiliation{Departamento de F\'{\i}sica At\'omica, Molecular y Nuclear,
and Instituto de F\'{\i}sica Te\'orica y Computacional Carlos I,
Universidad de Granada, Granada 18071, Spain}

\begin{abstract}
We study nuclear medium effects in Drell-Yan processes using quark parton distribution functions in a microscopic nuclear model which takes into account the effect of Fermi
 motion, nuclear binding and nucleon correlations through a relativistic nucleon spectral function. The contributions of $\pi$ and $\rho$ mesons are also included. The beam energy loss is calculated assuming 
 forward propagation of beam partons using Eikonal approximation. The results are compared with the theoretical and experimental results. The model is able to
successfully explain the low target $x_t$ results of E772 and E866 Drell-Yan experiments and is applicable to the forthcoming experimental analysis of E906 Sea Quest experiment at Fermi Lab.
\end{abstract}
\maketitle
\section{Introduction} 
Drell-Yan production of lepton pairs~\cite{drellyan} from nucleons and nuclear targets is an important tool to study the quark structure of nucleons and its modification in the nuclear
medium. In particular, the proton induced Drell-Yan production of muon pairs on nucleons and nuclei provides a direct probe to investigate the quark parton
distribution functions(PDFs). The Drell-Yan(DY) production takes place through basic process of
quark-antiquark annihilation into lepton pairs i.e $q^{b(t)}+\bar q^{t(b)}\rightarrow l^++l^-$ where b and t indicate the beam proton and the target nucleon. In this basic process a quark(antiquark) 
in the beam carrying a longitudinal momentum fraction $x_b$  interacts with an antiquark(quark) in the target carrying longitudinal momentum fraction $x_t$ of the target
momentum per nucleon to produce a virtual photon which decays into lepton pairs. 

The cross section per target nucleon $\frac{d^{2}\sigma}{dx_bdx_t}$ in the leading order is given by~\cite{Amanda}:
\begin{eqnarray}\label{DY1}
\frac{d^2\sigma}{dx_bdx_t}&=&\frac{4\pi\alpha^2}{9 Q^2}\sum_fe_f^{2}\left\{q_{f}^{b}(x_b,Q^2)\bar q_{f}^{t}(x_t,Q^2) + \bar q_{f}^{b}(x_b,Q^2)q_{f}^{t}(x_t,Q^2)\right\}
\end{eqnarray}
where $\alpha$ is the fine structure constant, $e_f$ is the charge of quark(antiquark) of flavor f, $Q^2$ is the photon virtuality and $q_{f}^{b,t}(x)$
and $\bar q_{f}^{b,t}(x)$ are the beam(target) quark/antiquark PDFs.

This process is directly sensitive to the antiquark parton distribution functions $\bar q(x)$ in target nuclei which has also 
 been studied by DIS experiments through the observation of EMC effect. Quantitatively the EMC effect describes the nuclear modification of nucleon structure function $F_2(x_t)$ for 
 the bound nucleon defined as $F_2(x_t)=x_t\sum_fe_f^2[q_f(x_t)~+~\bar q_f(x_t)]$ and gives information about the modification of the sum of quark and antiquark PDFs~\cite{kenyon,Geesaman:1995yd} which is
  dominated by the valence quarks in the high $x_t$ region ($x_t > 0.3$). In the low $x_t$ region ($x_t \le 0.3$), where sea quarks are expected to give dominant contribution, the study of $F_2(x)$ gives
   information about sea quark and antiquark PDFs. Thus, nuclear modifications are phenomenologically incorporated in $q(x_t)$ and $\bar q(x_t)$ using the experimental data on $F_2(x_t)$ and are used to analyze the DY yields from nuclear targets. Some authors succeed in giving a 
   satisfactory description of DIS and DY data on nuclear targets using same set of nuclear $q(x)$ and $\bar q(x)$~\cite{eskola09}, while some others find it difficult to provide a consistent description
   of DIS and DY data using the same set of nuclear PDFs~\cite{scheinbein2008}. On the other hand, there are many theoretical attempts to describe the nuclear modifications of quark and antiquark
    PDFs to explain DIS which have been used to describe the DY process on nuclear targets
    ~\cite{miller}-\cite{Marco:1997xb}. The known nuclear modifications discussed in literature in the case of DIS are (a) modification of nucleon structure inside the nuclear medium, (b) a significantly enhanced 
    contribution of subnucleonic degrees of freedom like pions or quark clusters in nuclei and (c) nuclear shadowing. 
    
      However, in the case of DY processes there is an additional nuclear effect due to initial state interaction of beam partons with the target partons which may be present before the hard collisions 
      of these partons produce lepton pairs. As the initial beam traverses the nuclear medium it loses energy due to interaction of beam partons with nuclear constituents of the target. This can be 
      visualized in terms of the
     interaction of hadrons or its constituents with the constituents of the target nucleus through various inelastic processes leading to energy loss of the interacting beam partons. This has been 
     studied phenomenologically using available parameterization of nuclear PDFs or theoretically in models based on QCD or Glauber approaches taking into account the 
     effect of shadowing which also plays an important role in the low $x_t$ region but any consensus in the understanding of physics behind the beam energy loss has been lacking~\cite{vasilev,Duan2005,Johnson2002,Garvey2003,Brodsky1993}. 
           This is also the region in which modification of sea quark PDF due to mesonic contributions are also important. It is however known that mesonic contributions enhance DY yields
           (and $F_2(x)$ in DIS) 
      while the shadowing and parton energy loss effects suppress them. Since, we are not studying the shadowing effect here in this work, therefore, 
      we confine ourselves to the region of $x_t > 0.1$, where shadowing does not play a major role. In this region, the main nuclear effects 
      are the mesonic contributions and nuclear structure effects as in the case of DIS with additional effect of parton energy loss in the beam parton energy due to the presence of nuclear targets. 
       
       In this paper, we present the results of nuclear medium effects on DY production
        of lepton pairs calculated in a theoretical microscopic nuclear model which has been successfully used to describe the DIS of charged leptons and $\nu$($\bar\nu$) from various nuclei
~\cite{marco1996,sajjadplb,sajjadnpa,prc84,prc85,prc87}. The model uses a relativistic nucleon spectral function to describe target nucleon momentum distribution incorporating Fermi motion, binding energy 
effects and nucleon correlations in a field theoretical model. The model has also been used to include the mesonic contributions from $\pi$ and $\rho$ mesons. The beam energy loss has been calculated in a model where the incoming beam proton loses energy in 
 inelastic collisions with the target hadrons as it travels the nuclear medium. This has been parameterized in terms of proton-nucleon scattering cross section using Glauber approach~\cite{Marco:1997xb}. The results have been presented for the 
kinematic region of experiments E772~\cite{alde} and E866~\cite{e866,vasilev} for proton induced DY processes in nuclear targets like $^{9}Be$, $^{12}C$, $^{40}Ca$, $^{56}Fe$ and $^{184}W$ in the region of $x_t > 0.1$. 
The numerical results extended up to $x_t=0.45$, should be useful in analyzing the forthcoming experimental results from the SeaQuest E906 experiment being done at Fermi Lab~\cite{Seaquest}.

     In section-\ref{sec:NE}, we present the formalism in brief, in section~\ref{sec:RD}, the results are presented and discussed and 
finally in section~\ref{sec:CC} we summarize the results and conclude our findings. 

\section{Nuclear effects}\label{sec:NE}
When DY processes take place in nuclei, nuclear effects appear which are generally believed to be due to \\
(a) nuclear structure arising from Fermi motion, binding energy and nucleon correlations\\
(b) additional contribution due to subnucleonic degrees of freedom like mesons and/or quark cluster in the nuclei and\\
(c) energy loss of the beam proton as it traverses the nuclear medium before producing lepton pairs.

In the case of proton induced DY processes in nuclei, the target nucleon has a Fermi momentum described by a momentum distribution. 
The target Bjorken variable $x_t$, defined for a free nucleon 
which is expressed 
covariantly as $x_t=\frac{2q.p_1}{(p_1+p_2)^2}$, where q is the four momentum of $\mu^+\mu^-$ pair, $p_{1\mu}$ and $p_{2\mu}$ are beam and target four momenta 
has a Fermi momentum dependence 
in the nuclear medium. Moreover, the projectile Bjorken variable $x_b$ expressed covariantly as $x_b=\frac{2 q.p_2}{(p_1+p_2)^2}$ also changes due to the energy loss of the beam particle caused by the  initial state interactions with the 
nuclear constituents as it travels through the nuclear medium before producing lepton pairs. These nuclear modifications in $x_b$ and $x_t$ are incorporated while evaluating Eq.~\ref{DY1}. Furthermore, there are 
additional contributions from the pion and rho mesons which are also taken into account. 

In the following, we briefly outline the model and refer to earlier work~\cite{Marco:1997xb,marco1996,sajjadplb,sajjadnpa,prc84,prc85,prc87} for details.              

\subsection{Nuclear Structure}
In a nucleus, scattering is assumed to take place from partons inside the individual nucleons which are bound in the nucleus and moving with a Fermi momentum $\vec p$. The target Bjorken variable 
$x_t$ becomes Fermi momentum dependent and PDF for quarks and antiquarks in the nucleus i.e. $q_f^t(x_t)$ and $\bar q_f^t(x_t)$ are calculated as a convolution of the PDFs 
in bound nucleon and a momentum distribution function of the nucleon inside the nucleus. The parameters of the momentum distribution are adjusted to correctly incorporate nuclear properties
 like binding energy, Fermi motion and the nucleon correlation effects in the nuclear medium. We use the Lehmann representation of the relativistic Dirac propagator for an interacting Fermi sea in nuclear 
 matter to derive such a momentum distribution and use Local Density Approximation to translate at a position $\bf r$ in the nucleus to describe the 
 finite nucleus~\cite{Marco:1997xb,sajjadnpa,prc84, prc85, prc87}. The relativistic propagator for a nucleon of mass $M_N$ is written in terms of 
 positive and negative energy components as
\begin{equation}  \label{prop2}
G^{0}(p_{0},{\bf p}) =\frac{M_N}{E({\bf p})}\left\{\frac{\sum_{r}u_{r}(p)\bar u_{r}(p)}{p^{0}-E({\bf p})+i\epsilon}+\frac{\sum_{r}v_{r}(-p)\bar v_{r}(-p)}{p^{0}+E({\bf p})-i\epsilon}\right\}
\end{equation} 
For a noninteracting Fermi sea where only positive energy solutions are considered the relevant propagator is rewritten in terms of occupation number $n({\bf p})=1$
 for p$\le p_{F}$ while $n({\bf p})$=0 for p$> p_{F}$:
\begin{eqnarray}  \label{prop4}
G^{0}(p_{0},{\bf p})&=&\frac{M_N}{E({\bf p})}\left\{\sum_{r}u_{r}(p)\bar u_{r}(p)\left[\frac{1-n(p)}{p^{0}-E({\bf p})+i\epsilon}+\frac{n(p)}{p^{0}-E({\bf p})-i\epsilon}\right]\right\}
\end{eqnarray}
 The nucleon propagator for a nucleon in an interacting Fermi sea is then calculated by making a perturbative expansion of $G(p_{0},{\bf p})$ in terms of $G^{0}(p_{0},{\bf p})$ given 
 in equation(\ref{prop2}) by retaining the positive energy contributions only (the negative energy components are suppressed).

This perturbative expansion is then summed in ladder approximation to give~\cite{marco1996,FernandezdeCordoba:1991wf}
\begin{eqnarray}
G(p_{0},{\bf p})&=&\frac{M_N}{E({\bf p})}\sum_{r}u_{r}(p)\bar u_{r}(p)\frac{1}{p^{0}-E({\bf p})}+\frac{M_N}{E({\bf p})}
\sum_{r}\frac{u_{r}(p)\bar u_{r}(p)}{p^{0}-E({\bf p})}\sum(p^{0},{\bf p})\frac{M_N}{E({\bf p})}\sum_{s}\frac{u_{s}(p)\bar u_{s}(p)}{p^{0}-E({\bf p})}+.....\\ \nonumber
&&=\frac{M_N}{E({\bf p})}\sum_{r}\frac{u_{r}(p)\bar u_{r}(p)}{p^{0}-E({\bf p})-\bar u_{r}(p)\sum(p^{0},{\bf p})u_{r}(p)\frac{M_N}{E({\bf p})}},
\end{eqnarray}
where $\sum(p^{0},{\bf p})$ is the nucleon self energy.

This allows us to write the relativistic nucleon propagator in a nuclear medium in terms of the Spectral functions of holes and 
particles as~\cite{FernandezdeCordoba:1991wf}
\begin{eqnarray}\label{Gp}
G (p^0, {\bf p})=\frac{M_N}{E({\bf p})} 
\sum_r u_r ({\bf p}) \bar{u}_r({\bf p})
\left[\int^{\mu}_{- \infty} d \, \omega 
\frac{S_h (\omega, \bf{p})}{p^0 - \omega - i \eta}
+ \int^{\infty}_{\mu} d \, \omega 
\frac{S_p (\omega, \bf{p})}{p^0 - \omega + i \eta}\right]\,
\end{eqnarray}
$S_h (\omega, \bf{p})$ and $S_p (\omega, \bf{p})$ being the hole
and particle spectral functions respectively, which are derived in Ref.~\cite{FernandezdeCordoba:1991wf}. 
We use:
\begin{equation}\label{sh}
S_h (p^0, {\bf p})=\frac{1}{\pi}\frac{\frac{M_N}{E({\bf p})}Im\Sigma^N(p^0,{\bf p})}{(p^0-E({\bf p})-\frac{M_N}{E({\bf p})}Re\Sigma^N(p^0,{\bf p}))^2
+ (\frac{M_N}{E({\bf p})}Im\Sigma^N(p^0,{\bf p}))^2}
\end{equation}
for $p^0 \le \mu$
\begin{equation}\label{sh1}
S_p (p^0, {\bf p})=-\frac{1}{\pi}\frac{\frac{M_N}{E({\bf p})}Im\Sigma^N(p^0,{\bf p})}{(p^0-E({\bf p})-\frac{M_N}{E({\bf p})}Re\Sigma^N(p^0,{\bf p}))^2
+ (\frac{M_N}{E({\bf p})}Im\Sigma^N(p^0,{\bf p}))^2}
\end{equation}
for $p^0 > \mu$. 

The normalization of this spectral function is obtained by imposing the baryon number conservation following the method of Frankfurt and Strikman~\cite{Frankfurt}.
 In the present paper, we use local density approximation (LDA) where we do not have a box of constant density, and the reaction takes 
place at a point ${\bf r}$, lying inside a volume element $d^3r$ with local density $\rho_{p}({\bf r})$ and $\rho_{n}({\bf r})$ 
corresponding to the proton and neutron densities at the point ${\bf r}$. This leads to the spectral functions for the protons and 
neutrons to be the function of local Fermi momentum given by
 \begin{eqnarray} \label{Fermi1}
k_{F_p}({\bf r})= \left[ 3\pi^{2} \rho_{p}({\bf r})\right]^{1/3}, k_{F_n}({\bf r})= \left[ 3\pi^{2} \rho_{n}({\bf r})\right]^{1/3}
\end{eqnarray}
and therefore the normalization condition may be imposed as
\begin{eqnarray} \label{norm2}
2\int\frac{d^{3}p}{(2\pi)^{3}}\int_{-\infty}^{\mu}S_{h}^{p(n)}(\omega,{\bf p},k_{F_{p,n}}({\bf r})) d\omega= \rho_{p,n}({\bf r})
\end{eqnarray}
 leading to the normalization condition given by
\begin{equation}\label{norm4}
2 \int d^3 r \;  \int \frac{d^3 p}{(2 \pi)^3} 
\int^{\mu}_{- \infty} \; S_h^{p(n)} (\omega, {\bf p}, \rho_{p(n)}({\bf r})) 
\; d \omega = Z(N)\,,
\end{equation}
where $\rho(r)$ is the baryon density for the nucleus which is normalized to A and is taken from the electron nucleus scattering experiments.
The average kinetic and total nucleon energy in a nucleus with the same number of protons and neutrons are given by: 
\begin{eqnarray}
<T>= \frac{4}{A} \int d^3 r \;  \int \frac{d^3 p}{(2 \pi)^3} (E({\bf p})-M_N) 
\int^{\mu}_{- \infty} \; S_h (p^0, {\bf p}, \rho(r)) 
\; d p^0\,,
\end{eqnarray}
\begin{eqnarray}
<E>= \frac{4}{A} \int d^3 r \;  \int \frac{d^3 p}{(2 \pi)^3}  
\int^{\mu}_{- \infty} \; S_h (p^0, {\bf p}, \rho(r)) 
\; p^0 d p^0\,,
\end{eqnarray}
and the binding energy per nucleon is given by~\cite{marco1996}:
\begin{equation}\label{be}
|E_A|=-\frac{1}{2}(<E-M_N>+\frac{A-2}{A-1}<T>)
\end{equation}
The binding energy per nucleon for each nucleus is correctly reproduced to match with the experimentally observed values. 
 This spectral function has been used to describe the DIS of charged leptons on the nuclear targets.
 In the case of nucleus, the nuclear hadronic tensor $W^{\mu \nu}_A$ for an isospin symmetric nucleus is derived to be~\cite{marco1996,sajjadnpa}:
\begin{eqnarray}\label{w2Anuclei}
W^{\mu \nu}_A&=& 2 \sum_{i=p,n} \int \, d^3 r \, \int \frac{d^3 p}{(2 \pi)^3} \, 
\frac{M_N}{E (\vec{p})} \, \int^{\mu}_{- \infty} d p^0 S_h (p^0, {\bf p},\rho_{i}) W^{\mu \nu}_i (p, q) 
\end{eqnarray}
where the factor 2 is a spin factor and using this the electromagnetic structure function $F_{2A}(x,Q^2)$ for a non-symmetric (N$\ne$Z) nucleus in DIS is obtained as~\cite{marco1996}, 
\begin{eqnarray}\label{f2Anuclei}
F^t_{2A}(x,Q^2)&=&2\sum_{i=p,n}\int d^3r\int\frac{d^3p}{(2\pi)^3}\frac{M_N}{E(\mathbf{p})}\int^{\mu_i}_{-\infty}dp^0\; 
S^{i}_{h}(p^0,\mathbf{p},\rho_i(r)) 
\sum_f e_f^2 x_t^\prime [q_f^i(x_t^\prime(p^0, {\vec p}))+\bar q_f^i(x_t^\prime(p^0, {\vec p}))] \nonumber\\
\end{eqnarray}
  For the numerical calculations, we have used CTEQ6.6~\cite{cteq} nucleon parton distribution functions(PDFs) for $q_f^i$ and $\bar q_f^i$. 
 $S_h^{i}$  are the two different spectral functions, each of them
normalized to the number of protons or neutrons in the nuclear target. $\rho_p(\rho_n)$ is the proton(neutron) density inside the nucleus.

We see that the nuclear structure effects like Fermi motion, binding energy and nucleon correlations are properly incorporated for bound quarks in nucleons in a nucleus and we write $q_{f}^t(x_t)$ as~\cite{Marco:1997xb}:
\begin{eqnarray}\label{qbarA}
q_{f}^t(x_t,Q^2)&=&2\sum_{i=p,n}\int d^3r\int\frac{d^3p}{(2\pi)^3}\frac{M_N}{E(\mathbf{p})} \int^{\mu_i}_{-\infty}dp^0\; 
S_{h}^i(p^0,\mathbf{p},\rho_i(r)) {q}_f^i(x_t^\prime(p^0, {\vec p}),Q^2)\nonumber\\
\bar q_{f}^t(x_t,Q^2)&=&2\sum_{i=p,n}\int d^3r\int\frac{d^3p}{(2\pi)^3}\frac{M_N}{E(\mathbf{p})} \int^{\mu_i}_{-\infty}dp^0\;
S_{h}^i(p^0,\mathbf{p},\rho_i(r)) {\bar q}_f^i(x_t^\prime(p^0, {\vec p}),Q^2),
\end{eqnarray}
 where ${q}_f^i(\bar q_{f}^i(x_t,Q^2))$ is the quark(antiquark) PDFs for flavor f inside a nucleon and the factor of 2 is because of quark(antiquark) spin degrees of freedom. ${x_t}^\prime = \frac{M_N}{p^0 - p_z} x_t$ which is obtained 
 from the covariant expression of $x_t^\prime=\frac{q\cdot p_1}{s_N}$ with $\vec p_1 \| z$ direction.
 
\subsection{Mesonic contributions}
Mesonic contributions are taken into account by making use of the imaginary part of the meson propagators instead of spectral function which were derived from the imaginary part of the
 propagator in the case of nucleon. So in the case of pion, we replace in Eq.(\ref{f2Anuclei})~\cite{sajjadnpa}
 \[\frac{M_N}{E(\mathbf{p})}\int^{\mu}_{-\infty}d\omega\; S_{h}(\omega,\mathbf{p})\; \delta(p^0-\omega)~\rightarrow~-\frac{1}{\pi}~\theta(p_0)\; ImD(p)\]
 where $D(p)$ is the pion propagator in the nuclear medium given by 
 \begin{equation}\label{dpi}
D (p) = [ {p^0}^2 - \vec{p}\,^{2} - m^2_{\pi} - \Pi_{\pi} (p^0, {\bf p}) ]^{- 1}\,,
\end{equation}
where
\begin{equation}\label{pionSelfenergy}
\Pi_\pi=\frac{f^2/m_\pi^2 F^2(p)\vec{p}\,^{2}\Pi^*}{1-f^2/m_\pi^2 V'_L\Pi^*}\,.
\end{equation}
Here, $F(p)=(\Lambda^2-m_\pi^2)/(\Lambda^2+\vec{p}\,^{2})$ is the $\pi NN$ form factor, $\Lambda$=1GeV, $f=1.01$, $V'_L$ is
the longitudinal part of the spin-isospin interaction and $\Pi^*$ is the irreducible pion self energy that contains the contribution of particle - hole and delta - hole excitations.

Following a similar procedure, as done in the case of nucleon, the contribution of the pions to hadronic tensor in the nuclear medium may be written as \cite{marco1996}
\begin{equation}\label{W2pion}
W^{\mu \nu}_{A, \pi} = 3 \int d^3 r \; \int \frac{d^4 p}{(2 \pi)^4} \;
\theta (p^0) (- 2) \; Im D (p) \; 2 m_\pi W^{\mu \nu}_{\pi} (p, q)
\end{equation}
However, Eq.(\ref{W2pion}) also contains the contribution of the pionic contents of the nucleon, which are already contained in the sea contribution of nucleon through Eq.(\ref{qbarA}), therefore, the pionic 
contribution of the nucleon is to be subtracted from Eq.(\ref{W2pion}), in order to calculate the contribution from the excess pions in the nuclear medium. This is obtained by replacing $I m D (p)$ by 
$\delta I m D (p)$~\cite{marco1996} as
\begin{equation}
Im D (p) \; \rightarrow \; \delta I m D (p) \equiv I m D (p) - \rho \;
\frac{\partial Im D (p)}{\partial \rho} \left|_{\rho = 0} \right.
\end{equation}

Using Eq.\ref{W2pion}, pion structure function $F_{2, \pi}^A(x_A)$ in a nucleus is derived as
\begin{equation}  \label{F2pion}
F_{2, \pi}^A (x) = - 6 \int  d^3 r  \int  \frac{d^4 p}{(2 \pi)^4} \; 
\theta (p^0) \; \delta I m D (p) \; 
\; \frac{x}{x_\pi} \; 2 M_N \; \sum_f e_f^2 x_\pi [q^f_{\pi}(x_\pi(p^0, {\vec p}))+\bar q^f_{\pi}(x_\pi(p^0, {\vec p}))]     \; \theta (x_\pi - x) \; 
\theta (1 - x_\pi), 
\end{equation}
where $\frac{x}{x_{\pi}} = \frac{- p^0 + p^z}{M_N}$.

Following a similar procedure, as done in the case of nucleon, the expression for the pion quark PDF in the nuclear medium $q_{f,\pi}^t(x_t,Q^2)$ is derived as~\cite{Marco:1997xb}:
\begin{eqnarray}  \label{F2piqbar}
q_{f,\pi}^t(x_t,Q^2) = - 6 \int  d^3 r  \int  \frac{d^4 p}{(2 \pi)^4} \theta (p^0) \delta I m D_\pi (p)  2 M_N  q_{f,\pi}(x_\pi)  \theta (x_\pi - x_t)  \theta (1 - x_\pi).
\end{eqnarray}
and a similar expression for $\bar q_{f,\pi}^t(x_t,Q^2)$. 

Similarly the contribution of the $\rho$-meson cloud to the structure function is taken into account
 in analogy with the above prescription and the rho structure function is written as~\cite{marco1996}
\begin{equation} \label{F2rho}
F_{2, \rho}^A(x) = - 12 \int d^3 r \int \frac{d^4 p}{(2 \pi)^4}
\theta (p^0) \delta Im D_{\rho} (p) \frac{x}{x_{\rho}} \, 2 M_N
\sum_f e_f^2 x_{\rho}[q_\rho^f(x_\rho(p^0, {\vec p}))+\bar q_\rho^f(x_{\rho}(p^0, {\vec p}))] \theta (1 - x_{\rho})\theta (x_\rho - x)
\end{equation}
and the expression for the rho PDF $q_{f,\rho}^t(x_t,Q^2)$ is derived as~\cite{Marco:1997xb}:
\begin{eqnarray}  \label{F2piqbar}
q_{f,\rho}^t(x_t,Q^2) = - 12 \int  d^3 r  \int  \frac{d^4 p}{(2 \pi)^4} \theta (p^0) \delta I m D_\rho (p)
  2 M_N  q_{f,\rho}(x_\rho)  \theta (x_\rho - x_t)  
\theta (1 - x_\rho),
\end{eqnarray}
\noindent
where $D_{\rho} (p)$ is now the $\rho$-meson propagator in the medium given by:
\begin{equation}\label{dro}
D_{\rho} (p) = [ {p^0}^2 - \vec{p}\,^{2} - m^2_{\pi} - \Pi^*_{\rho} (p^0, {\bf p}) ]^{- 1}\,,
\end{equation}
where
\begin{equation}\label{pionSelfenergy}
\Pi^*_\rho=\frac{f^2/m_\rho^2 C_\rho F_\rho^2(p)\vec{p}\,^{2}\Pi^*}{1-f^2/m_\rho^2 V'_T\Pi^*}\,.
\end{equation}
Here, $V'_T$ is the transverse part of the spin-isospin interaction, $C_\rho=3.94$, $F_\rho(p)=(\Lambda_\rho^2-m_\rho^2)/(\Lambda_\rho^2+\vec{p}\,^{2})$ is the $\rho NN$ form factor, 
$\Lambda_\rho$=1GeV, $f=1.01$, and $\Pi^*$ is the irreducible rho self energy that contains the contribution of particle - hole and delta - hole excitations, 
$\frac{x}{x_{\rho}} = \frac{- p^0 + p^z}{M_N}$. Quark and antiquark PDFs for pions have been taken from the parameterization given by Gluck et al. Ref.\cite{Gluck:1991ey} and for the 
rho mesons we have taken the same PDFs as for the pions. 

\subsection{Energy loss of beam partons}
The incident proton beam traverses the nuclear medium before the beam parton undergoes a hard collision with the target parton. The incident proton may lose energy due to soft inelastic collisions, it 
 might scatter on its way within the nucleus before producing a lepton pair. We shall consider the region $x_t > 0.1$ (away from the shadowing region), 
 and assume that the initial state interactions are manifested through the inelastic proton-proton collisions in the case of 
 proton induced DY processes from nuclei. We further assume that each collision of this type occurs with a probability $\sigma_N \rho dl$ during a length of l in the nuclear medium with nuclear density 
 $\rho$, the proton loses a fraction $\beta$ of its energy. In principle, $\beta$ may depend upon energy but we assume it to be constant. With this assumption, the energy $E_b$ of the beam parton 
 is described by~\cite{Marco:1997xb}: 
 \begin{eqnarray}\label{eloss1}
\frac{d E_b}{dl} &=& - \sigma_{NN} \rho \beta E_b\nonumber\\
E_b (\vec{r}) &=& E_{b \, in} \, \exp [- \beta \sigma_{NN} \int_{- \infty}^z
\rho (\vec{b} , z') dz']
\end{eqnarray}
Since $x_b$ is inversely proportional to $E_b$, we write
\begin{eqnarray}\label{eloss}
x_b (\vec{r}) &=& x_b \, \exp [\beta \sigma_{NN} \int_{-\infty}^z \rho
(\vec{b}, z') d z']
\end{eqnarray}

where $\sigma_{NN}$ is the $NN$ total cross section,
($\sigma_{NN}$) taken to be 40mb~\cite{pdg}, $\rho (\vec{r})$
the nuclear density and $\vec{b}$ the impact parameter. 

Using these modified values of $x_b(\bf r)$, the DY cross sections are written as
 \begin{eqnarray}\label{spec}
 \frac{d^2 \sigma^{(N)}}{d x_b dx_t} &=& \frac{4 \pi \alpha^2}{9 q^2} 4 \int d^3 r 
\sum_f e_f^2 \left[q_{f,p} (x_b (\vec{r})) \int \frac{d^3 p}{(2 \pi)^3} \frac{M_N}{E (\vec{p})}
\int_{- \infty}^\mu d p^0 S_h (p^0, {\bf p}) \bar{q}_{f,N} (x_t^\prime)\right.\nonumber\\
 &+& \bar{q}_{f,p} (x_b (\vec{r})) 
 \left.\int \frac{d^3 p}{(2 \pi)^3} \frac{M_N}{E
(\vec{p})} \int_{ - \infty}^\mu d p^0 S_h (p^0, {\bf p}) 
q_{f,N} (x_t^\prime)\right] \theta (x_t^\prime) \theta (1 - x_t^\prime)~\theta (1 - x_b (\vec{r}))
\end{eqnarray}
where $S_h (p^0, {\bf p})$ is the hole spectral function for the nucleon in the nucleus. $q_{f,N}=\frac{1}{2}(q_{f,p}+q_{f,n})$ and
 $\bar q_{f,N}=\frac{1}{2}(\bar q_{f,p}+ \bar q_{f,n})$ are the nucleon PDFs of flavor f averaged over quark and antiquark PDFs in proton and neutron.
\begin{figure}
\includegraphics[scale=0.5]{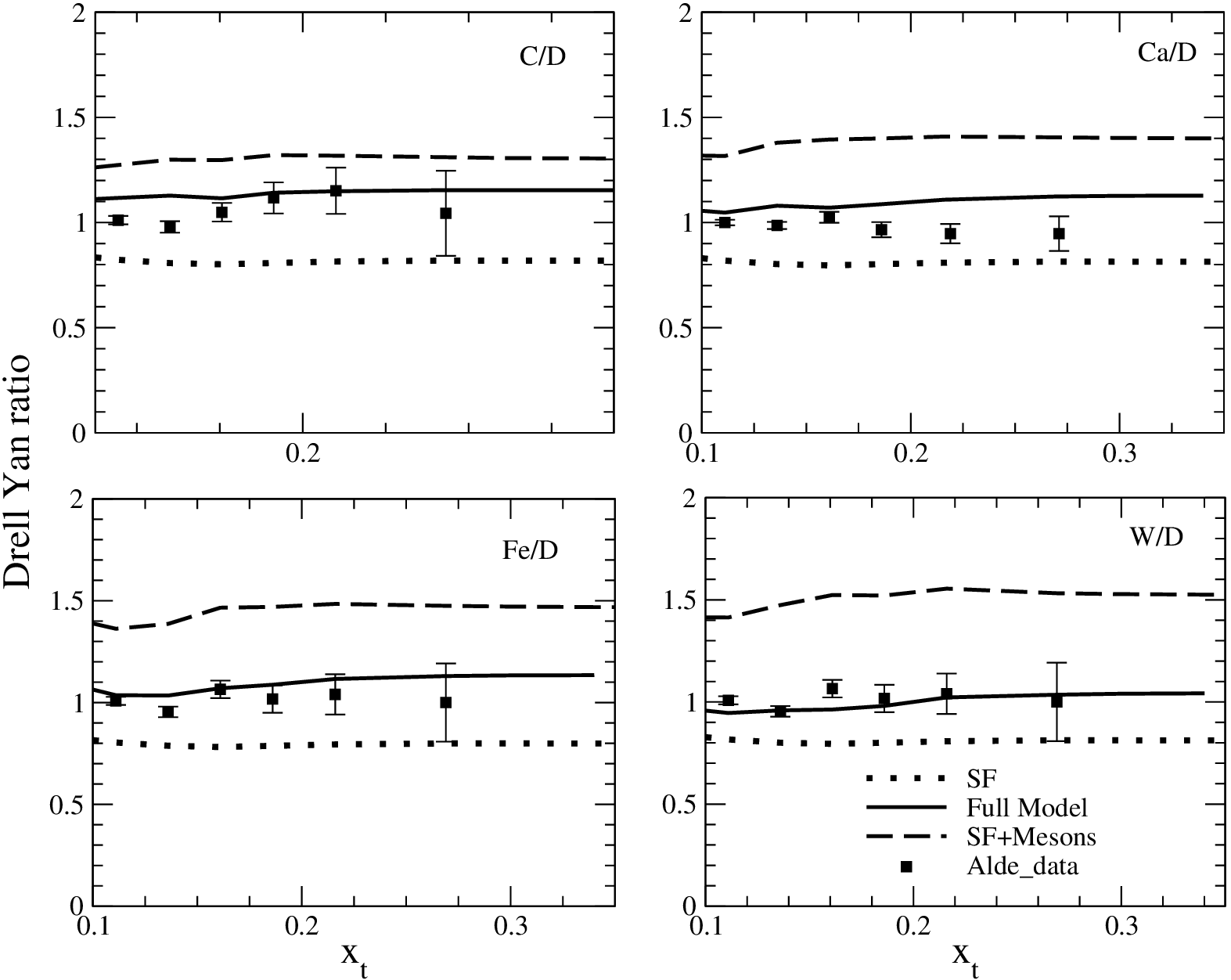}
\caption{Left panel: $\frac{\frac{d\sigma}{dx_t}(C, Fe)}{\frac{d\sigma}{dx_t}(D)}$ vs $x_t$ at $\sqrt{s_N}$=38.8GeV. 
 Spectral function without energy loss: dotted line, spectral function+meson cloud contributions without energy loss:dashed line and the solid line: results with the full model i.e. 
 spectral function+meson cloud contributions with energy loss. For the energy loss we have taken $\beta=0.04$. 
Experimental points are of E772 experiment~\cite{alde}. Right panel: 
$\frac{\frac{d\sigma}{dx_t}(Ca, W)}{\frac{d\sigma}{dx_t}(D)}$ vs $x_t$, lines have same meaning as in the left panel.}
\label{fig1}  
\end{figure}
The pion and rho cloud contributions are written as~\cite{Marco:1997xb}:
\begin{eqnarray}\label{spec1}
 \frac{d^2 \sigma^{(\pi)}}{d x_b d x_t}
&=& \frac{4 \pi \alpha^2}{9 q^2} (-6) \int d^3 r \sum_f e_f^2 \left[q_{f,p} (x_b (\vec{r})) \int \frac{d^4 p}{(2 \pi)^4}
\theta (p^0) \delta Im D_\pi (q) 2 M_N \bar{q}_{f,\pi} (x_{\pi})\right.\nonumber\\
 &+& \left.\bar{q}_{f,p} (x_b (\vec{r})) \int \frac{d^4 p}{(2 \pi)^4} \theta (p^0)
\delta \, Im \, D_\pi (q) 2 M_N q_{f,\pi} (x_{\pi})\right] \theta (x_{\pi} - x_t) \; \theta (1 - x_{ \pi}) \; 
\theta (1 - x_b (\vec{r}))
\end{eqnarray}
and 
\begin{eqnarray}\label{specrho}
 \frac{d^2 \sigma^{(\rho)}}{d x_b d x_t}
&=& \frac{4 \pi \alpha^2}{9 q^2} (-12) \int d^3 r \sum_f e_f^2 \left[q_{f,p} (x_b (\vec{r})) \int \frac{d^4 p}{(2 \pi)^4}
\theta (p^0) \delta Im D_\rho (q) 2 M_N \bar{q}_{f,\rho} (x_{ \rho})\right.\nonumber\\
 &+& \left.\bar{q}_{f,p} (x_b (\vec{r})) \int \frac{d^4 p}{(2 \pi)^4} \theta (p^0)
\delta \, Im \, D_\rho (q) 2 M_N q_{f,\rho} (x_{\rho})\right] \theta (x_{ \rho} - x_t) \; \theta (1 - x_{ \rho}) \; 
\theta (1 - x_b (\vec{r}))
\end{eqnarray}
Using Jacobian transformation Eq. \ref{spec} may be written as:
\begin{eqnarray}\label{duan1}
\frac{d^2\sigma}{dx_bdM}&=&\frac{8\pi\alpha^2}{9 M}\frac{1}{x_b s_N} 4 \int d^3 r 
\sum_f e_f^2 \left[q_{f,p} (x_b (\vec{r})) \int \frac{d^3 p}{(2 \pi)^3} \frac{M_N}{E (\vec{p})}
\int_{- \infty}^\mu d p^0 S_h (p^0, {\bf p}) \bar{q}_{f,N} (x_t^\prime)\right.\nonumber\\
&+& \bar{q}_{f,p} (x_b (\vec{r})) 
 \left.\int \frac{d^3 p}{(2 \pi)^3} \frac{M_N}{E
(\vec{p})} \int_{ - \infty}^\mu d p^0 S_h (p^0, {\bf p}) 
q_{f,N} (x_t^\prime)\right] \theta (x_t^\prime) \theta (1 - x_t^\prime)~\theta (1 - x_b (\vec{r}))
\end{eqnarray}
\begin{figure}
\includegraphics[scale=0.5]{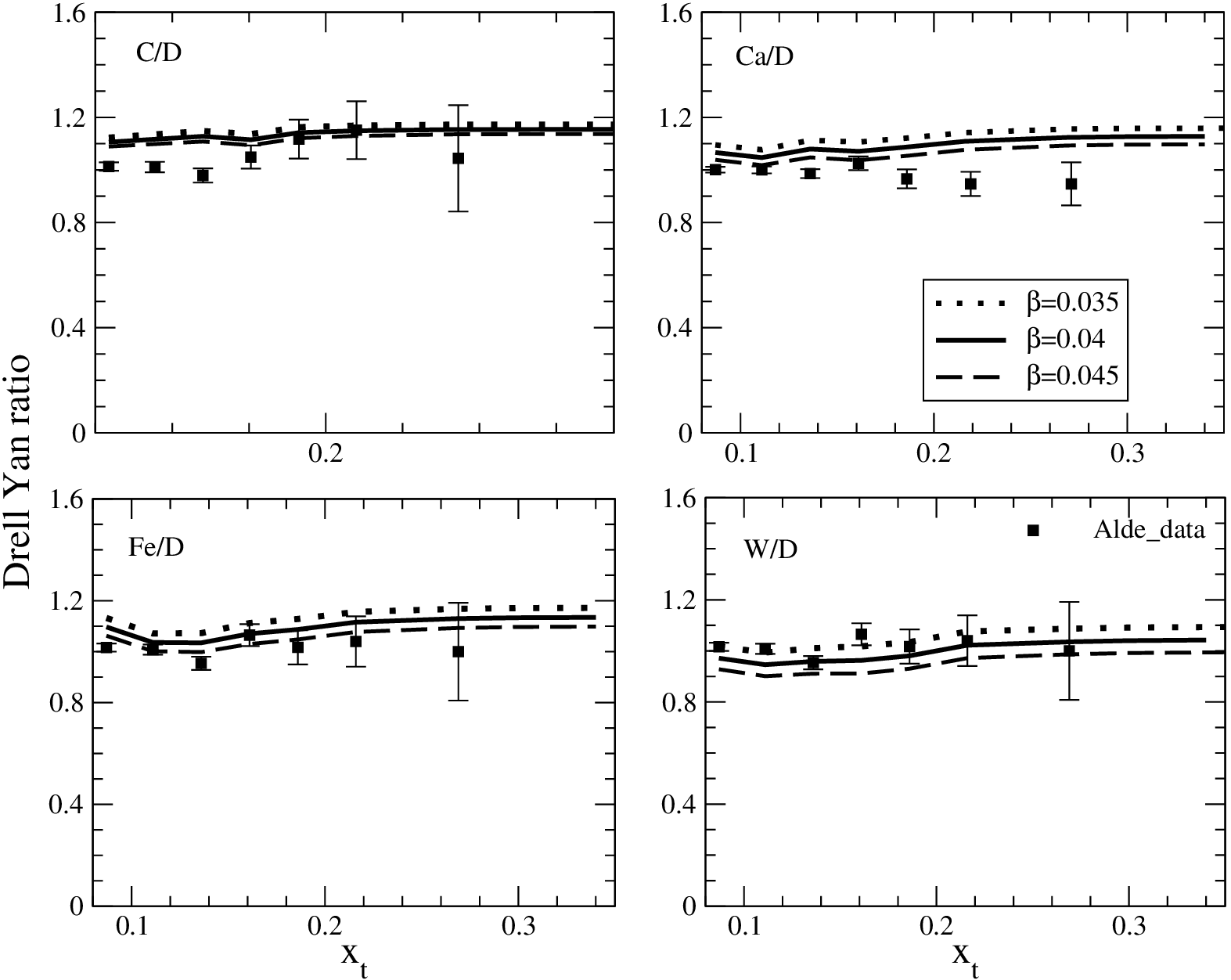}
\caption{$\frac{\frac{d\sigma}{dx_t}(i)}{\frac{d\sigma}{dx_t}(D)}$ vs $x_t$ at $\sqrt{s_N}$=38.8GeV for $\beta=0.04$. i stands for the various nuclei like C, Ca, Fe and W. Results are shown for the full 
model(spectral function+meson cloud contribution) with 
different values of $\beta$, a parameter used in the expression of the energy loss. 
Experimental points are of E772 experiment~\cite{alde}.}
\label{fig2}  
\end{figure}

\begin{figure}
\includegraphics[scale=0.5]{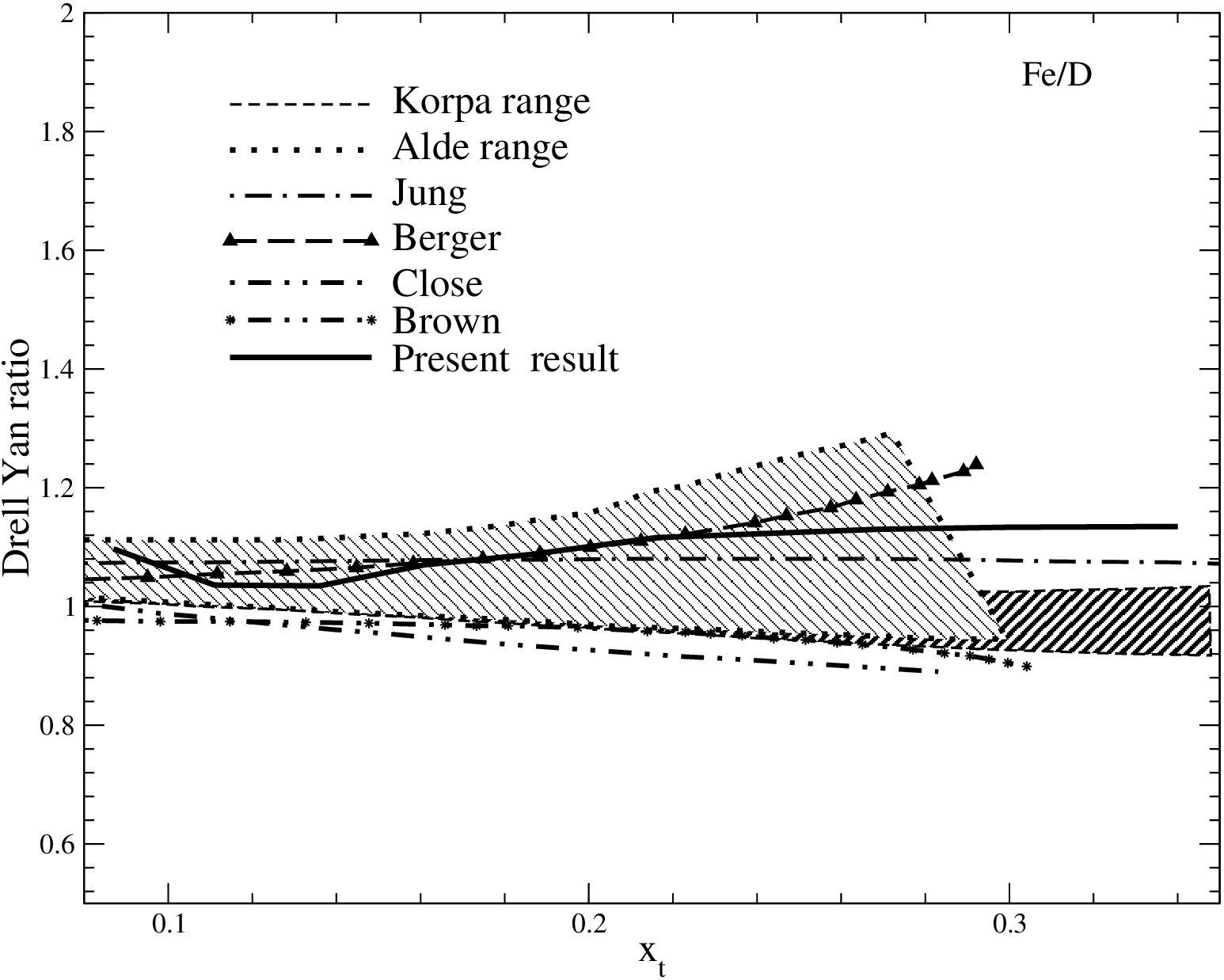}
\caption{$\frac{\frac{d\sigma}{dx_t}(Fe)}{\frac{d\sigma}{dx_t}(D)}$ vs $x_t$ at $\sqrt{s_N}$=38.8GeV. These results are shown for
$\beta=0.04$. Theoretical results of Korpa et al. ~\cite{Dieperink:1997iv} using different parameter values and Alde et al.~\cite{alde}
in the different models are shown through bands.
Dashed line with triangle up: Berger and Coester~\cite{Berger:1984na} results, dotted dashed line: Jung and Miller~\cite{Jung:1990pu}
results, double dotted dashed line: results of Close et al.~\cite{Close:1984zn} 
and Double dotted with stars: results of Brown et al.~\cite{Brown:1993sua}. Solid line is our result with spectral function and meson 
cloud contributions.}
\label{fig3}  
\end{figure}

 \begin{figure}
\includegraphics[scale=0.5]{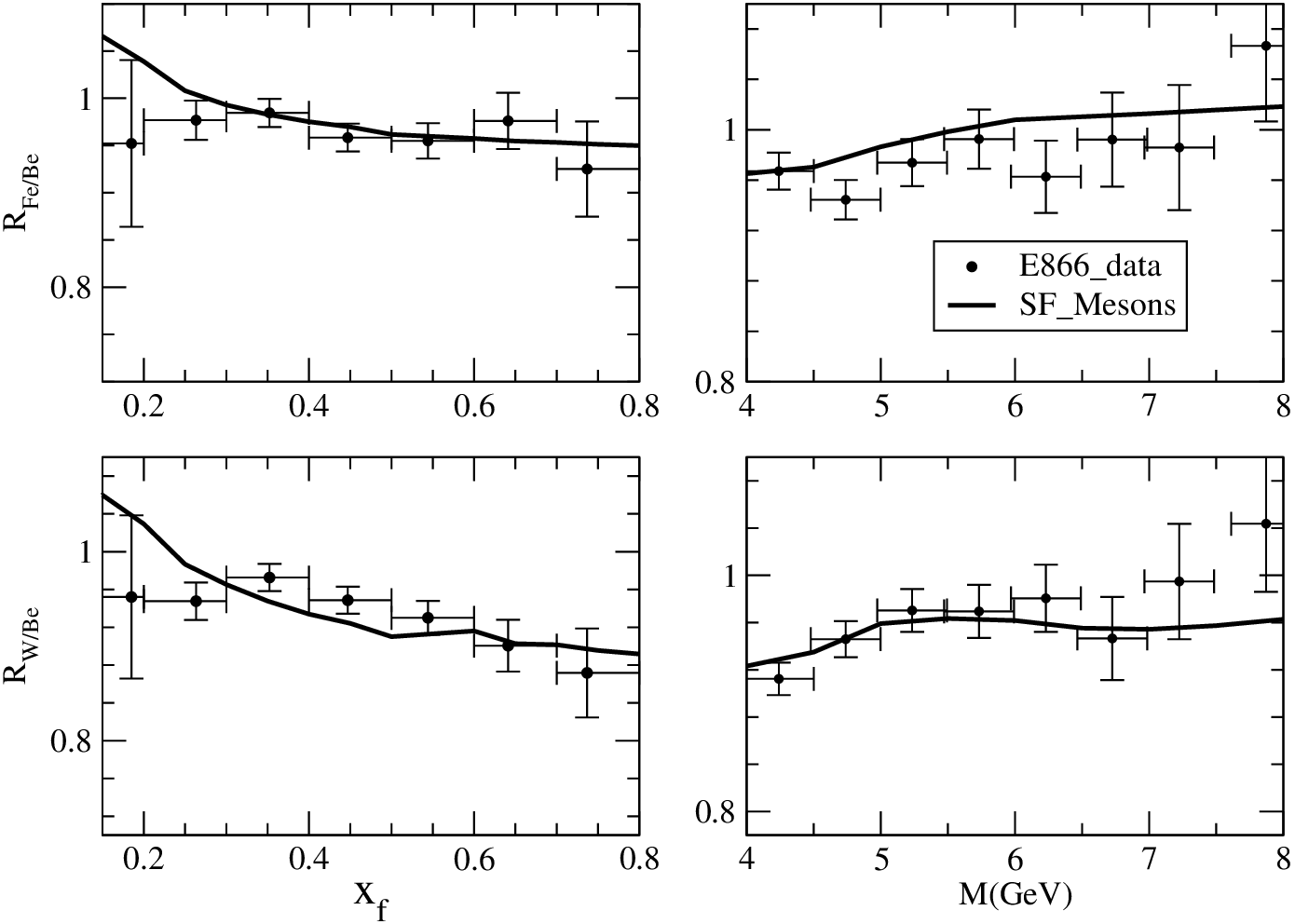}
\caption{ Left Panel: $\frac{\frac{d\sigma}{dx_F}(Fe,W)}{\frac{d\sigma}{dx_F}(Be)}$ vs $x_F$ for $\beta=0.04$, Right Panel: $\frac{\frac{d\sigma}{dM}(Fe,W)}{\frac{d\sigma}{dM}(Be)}$ vs $M(=\sqrt{x_bx_ts_N})$GeV, 
at $\sqrt{s_N}$=38.8GeV. Experimental points are of E866 experiment~\cite{e866,vasilev}. 
Spectral Function+Meson cloud contribution: solid line.}\label{fig4}  
\end{figure}

To evaluate proton-deuteron Drell-Yan cross section, we write 
\begin{eqnarray}\label{F2A_Deut1}
\frac{d\sigma^{pd}}{dx_b dx_t} 
&=& 
\frac{d\sigma^{pp}}{dx_b dx_t} + \frac{d\sigma^{pn}}{dx_b dx_t}\,.
\label{eq:dy-pd-exp}
\end{eqnarray}
To take into account the deuteron effect, the quark/antiquark distribution function inside the deuteron target have been
calculated using the same formula as for the nuclear structure function but performing the convolution with the deuteron 
wave function squared instead of using the spectral function with the Paris N-N potential. 

In terms of the deuteron wave function,  one may write
\begin{eqnarray}\label{F2A_Deut}
q_{f}^t(x_t,Q^2)=\int \frac{d^3p}{(2\pi)^3}|\Psi_D(\mathbf{p})|^2 {q}_f^{N}(x_t^\prime({\bf p}),Q^2)
\end{eqnarray}
and similar expression for the antiquarks.

\section{Results and Discussion}\label{sec:RD}
\begin{figure}
\includegraphics[scale=0.5]{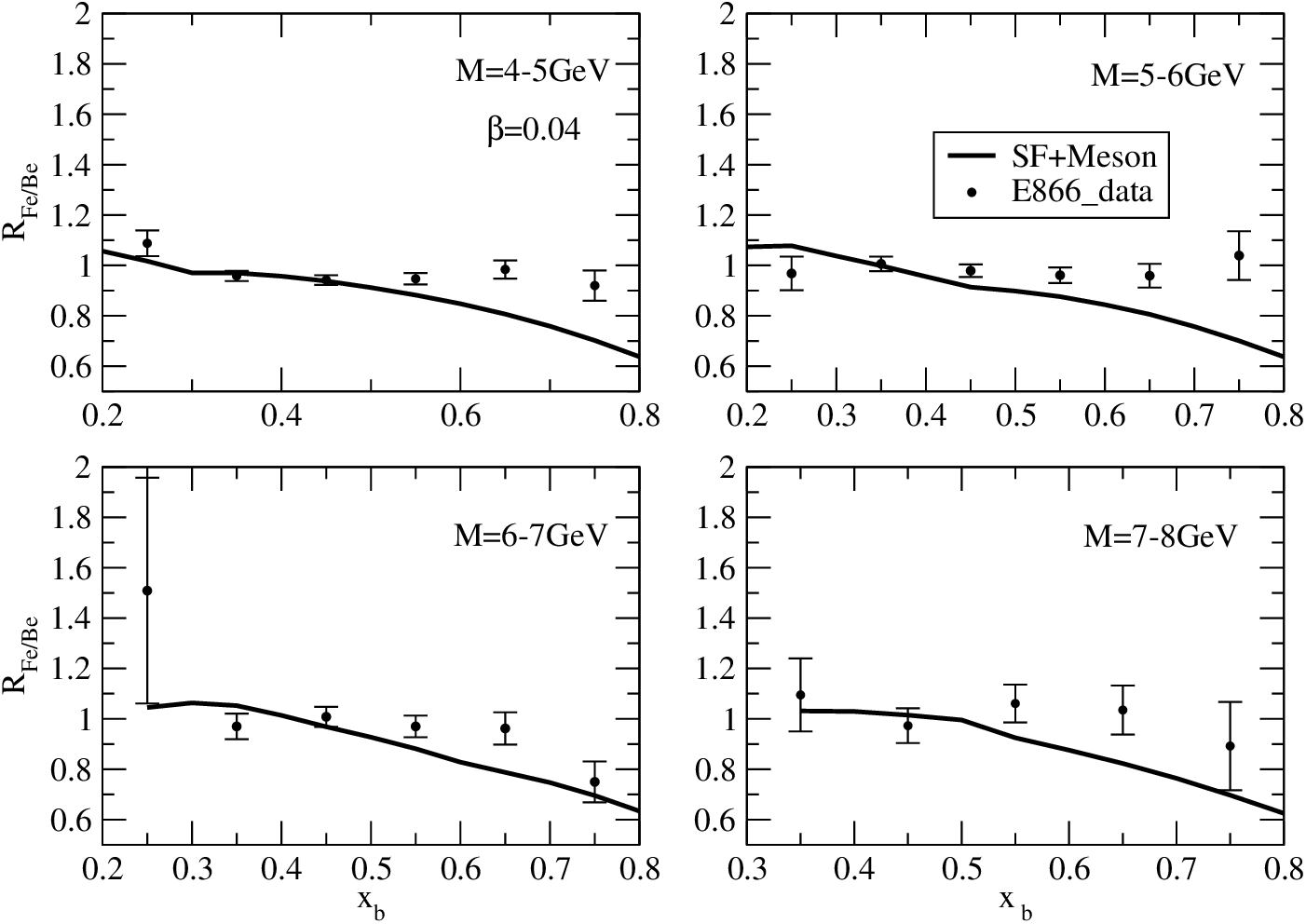}
\caption{$\frac{\frac{d\sigma}{dM dx_b}(Fe)}{\frac{d\sigma}{dM dx_b}(Be)}$ vs $x_b$ at different $M(=\sqrt{x_bx_ts_N})$ with $\sqrt{s_N}$=38.8GeV. 
Experimental points are of E866 experiment~\cite{e866,vasilev}.  Spectral Function+Meson cloud contribution: solid line.}\label{fig5}  
\end{figure}
 The numerical results for $\left(\frac{d\sigma}{dx_t}\right)^A$ for a nucleus A have been evaluated after integrating $\left(\frac{d\sigma}{dx_b dx_t}\right)^A=\left(\left(\frac{d\sigma}{dx_b dx_t}\right)^N
 ~+~\left(\frac{d\sigma}{dx_b dx_t}\right)^\pi~+~\left(\frac{d\sigma}{dx_b dx_t}\right)^\rho\right)$ over $x_b$ from $x_b=x_t+0.26$ to $x_b= 1$ for $Q^2~>~16GeV^2$. The cross sections 
 $\left(\frac{d\sigma}{dx_b dx_t}\right)^{i=N,\pi,\rho}$ are evaluated using Eqns.\ref{spec}, \ref{spec1} and \ref{specrho} respectively, where the nucleon quark(antiquark) PDFs given by 
 CTEQ6.6~\cite{cteq} and pion quark(antiquark) PDFs given by  Gluck et al.~\cite{Gluck:1991ey} have been used. The spectral function $S_h (p^0, {\bf p})$
 with parameters fixed by  Eqns.(\ref{norm2}),~(\ref{norm4}) 
 has been used to calculate the nucleon contribution which reproduce the binding energy per nucleon given in Eq.~\ref{be} and has no free parameter. For evaluating the mesonic contributions Eqns.(\ref{spec1}) and (\ref{specrho}) 
 have been used with the parameters of $D_\pi$ and $D_\rho$ fixed so that the experimental data on $F_2^i(x_t)$ for various nuclei $i=^{9}Be, ~^{12}C, ~^{40}Ca ~{\rm and}~ ^{56}Fe$ are 
 reproduced satisfactorily~\cite{sajjadnpa}. 
 Taking the energy loss parameter $\beta$ in Eq.(\ref{eloss}) as a variable parameter, we present our results in Fig. \ref{fig1}
 for $\frac{\left(\frac{d\sigma}{dx_t}\right)^{i}}{\left(\frac{d\sigma}{dx_t}\right)^D}$ 
 using $\beta=0.04$ and compare them with the experimental results of E772~\cite{alde} for i=$^{12}C$, $^{40}Ca$, $^{56}Fe$, and $^{184}W$ nuclei. 
 In the numerical evaluation of the denominator $\left(\frac{d\sigma}{dx_t}\right)^D$ i.e. Drell-Yan cross section for the proton-deuteron scattering, we have obtained the
 results by using Eq.~\ref{F2A_Deut1} and Eq.~\ref{F2A_Deut} (with and without the deuteron effect), where Eq.~\ref{F2A_Deut} takes care of deuteron effect. 
 We find the deuteron effect to be small on the ratio R (about 2$\%$) and have not been shown in this figure.
 We find that the nuclear structure effects due to bound nucleon lead to a suppression in the DY yield of about  $16-18\%$
 in the region of $0.1~<~x_t~<0.3$ which is larger than what has been found in the case of $F_2(x_t)$~\cite{sajjadnpa}.
 On the other hand, there is significant contribution of mesons which increases the DY ratio and overestimates the DY yields 
 which increases with A. For example, in the case of $^{12}C$ it is around $18\%$, $25\%$ in $^{40}Ca$, $35\%$ in $^{56}Fe$ and $45\%$ in $^{184}W$
  in this range of $x_t$. This increase in the DY yield from meson cloud contribution increases with A.
  Thus, the mesonic contribution found in the case of DY yields is larger than found in the case
  of $F_2(x_t)$ for the nuclei studied here and in Ref.~\cite{sajjadnpa}. This contribution is also found
  to be sensitive to the parameters used in the meson propagators $D_\pi(p)$ and $D_\rho(p)$, but we have used the same parameters which satisfactorily produce the results for the electromagnetic 
  structure function $F_2(x)$ for all the nuclei like $^{9}Be$, $^{12}C$, $^{40}Ca$ and $^{56}Fe$ and do not treat them as free parameters.
   We find the contribution from rho meson cloud to be much smaller than the contribution from pion cloud. 
  When we include the energy loss effect, we find that there is a suppression in the DY yield which further decreases with the increase in mass number A.
  For example, in the case of $^{12}C$ it is around $12\%$, $20\%$ in $^{40}Ca$, $25\%$ in $^{56}Fe$, and $35\%$ in $^{184}W$. 
  The increase in DY yield due to mesonic contribution and suppression due to beam energy loss compensate each other and we get 
  a reasonable agreement with the experimental results for $\beta=0.04$. However, the numerical value of beta needed to reproduce the experimental results
  can vary depending  upon the value of parameter Lambda($\Lambda_\pi$ or $\Lambda_\rho$) used for evaluating the propagators $D_\pi$ and $D_\rho$ 
  in Eqns. \ref{dpi} and \ref{dro}. A smaller value of beta$(<0.04)$ would also reproduce the experimental results provided a smaller value of parameter Lambda $(<1 GeV)$ is used to evaluate $D_\pi$ and $D_\rho$.
 In Fig. 2, we show the dependence of DY yield ratio on the energy loss parameter $\beta$ for Lambda =1GeV.
  
 In Fig.\ref{fig3}, we present our results for $\frac{\left(\frac{d\sigma}{dx_t}\right)^{Fe}}{\left(\frac{d\sigma}{dx_t}\right)^D}$, with $x_b = x_t+0.26$, and compare the results with the results of 
 various theoretical calculations available in the literature~\cite{Dieperink:1997iv,alde,Jung:1990pu,Berger:1984na,Close:1984zn,Brown:1993sua}.  Our results are 
presented for the full model with $\beta=0.04$. We see from Fig.\ref{fig3}, that our 
results agree with the results of Close et al.~\cite{Close:1983tn}, Berger et al.~\cite{Berger:1984na} and Jung and Miller~\cite{Jung:1990pu}. 
\begin{figure}
\includegraphics[scale=0.5]{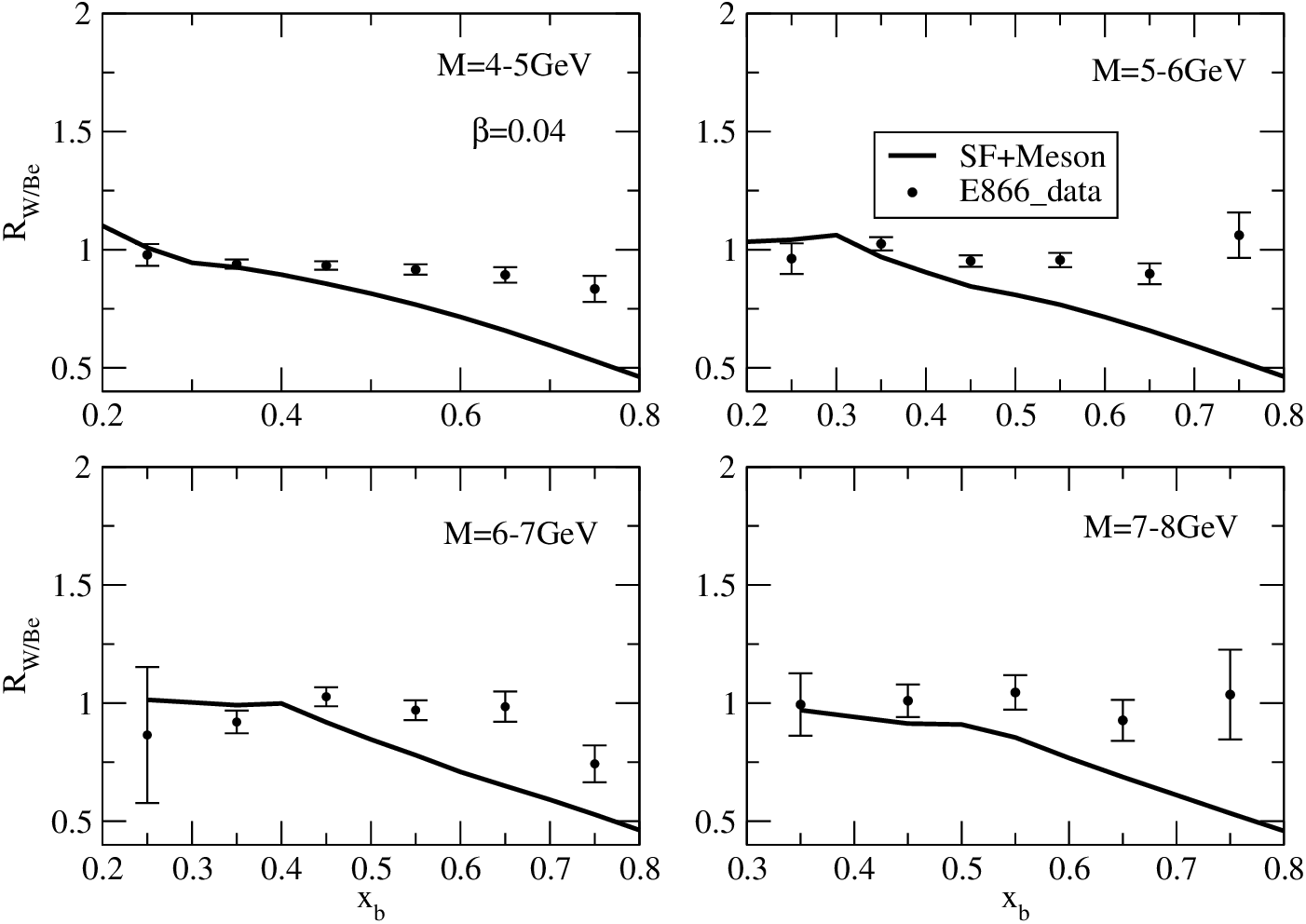}
\caption{$\frac{\frac{d\sigma}{dM dx_b}(W)}{\frac{d\sigma}{dM dx_b}(Be)}$ vs $x_b$ at different $M(=\sqrt{x_bx_ts_N})$ with $\sqrt{s_N}$=38.8GeV. 
Experimental points are of E866 experiment~\cite{e866,vasilev}.  Spectral Function+Meson cloud contribution: solid line.}\label{fig6}  
\end{figure}

In E866 experiment~\cite{e866,vasilev} the results are also presented for $\frac{d\sigma}{dx_F}$ vs $x_F$, where $x_F=x_b - x_t$ and  $\frac{d\sigma}{dM}$ vs $M$, where $M(=\sqrt{x_bx_ts_N})$. 
 Using Eqns.\ref{spec} and \ref{duan1} we have obtained the results respectively for $\frac{d\sigma}{dx_F}$ vs $x_F$ and  $\frac{d\sigma}{dM}$ vs $M$ and shown these results in 
 Fig.\ref{fig4}. For $\frac{d\sigma}{dx_F}$ vs $x_F$, 
 we have integrated over $x_b$ between the limits $0.21 \le x_b \le 0.95$ and following the kinematical cuts of $4.0<M<8.4$ GeV used in E866~\cite{e866,vasilev} experiment. In the case of $\frac{d\sigma}{dM}$ vs $M$, we have integrated over $x_b$ between the limits $0.21 \le x_b \le 0.95$ and put the
 kinematical constraint $0.13 \le x_F \le 0.93$ as used in E866~\cite{e866,vasilev} experiment. 
 These results are shown for the Drell-Yan ratio for $\frac{\left(\frac{d\sigma}{dx_F}\right)^{i}}{\left(\frac{d\sigma}{dx_F}\right)^{Be}}$ vs $x_F$(Left panel) and 
$\frac{\left(\frac{d\sigma}{dM}\right)^{i}}{\left(\frac{d\sigma}{dM}\right)^{Be}}$ vs $M$(Right panel), i stands for the iron nucleus(top panel) and tungsten nucleus(bottom panel).  
We find a good agreement with the experimental results for the various Drell-Yan ratios available from E866~\cite{e866,vasilev} experiment. 

In Fig. \ref{fig5}, we present the results for the Drell-Yan ratio for $\frac{\left(\frac{d^2\sigma}{dx_bdM}\right)^{Fe}}{\left(\frac{d^2\sigma}{dx_bdM}\right)^{Be}}$ vs $x_b$(Left panel) for different values
 of $M(=\sqrt{x_bx_ts_N})$, between the same kinematic limits as taken in the numerical evaluation of the results for Fig.\ref{fig4}. The results for this ratio for tungsten to beryllium target are shown in  Fig. \ref{fig6}.

\begin{figure}
\includegraphics[scale=0.5]{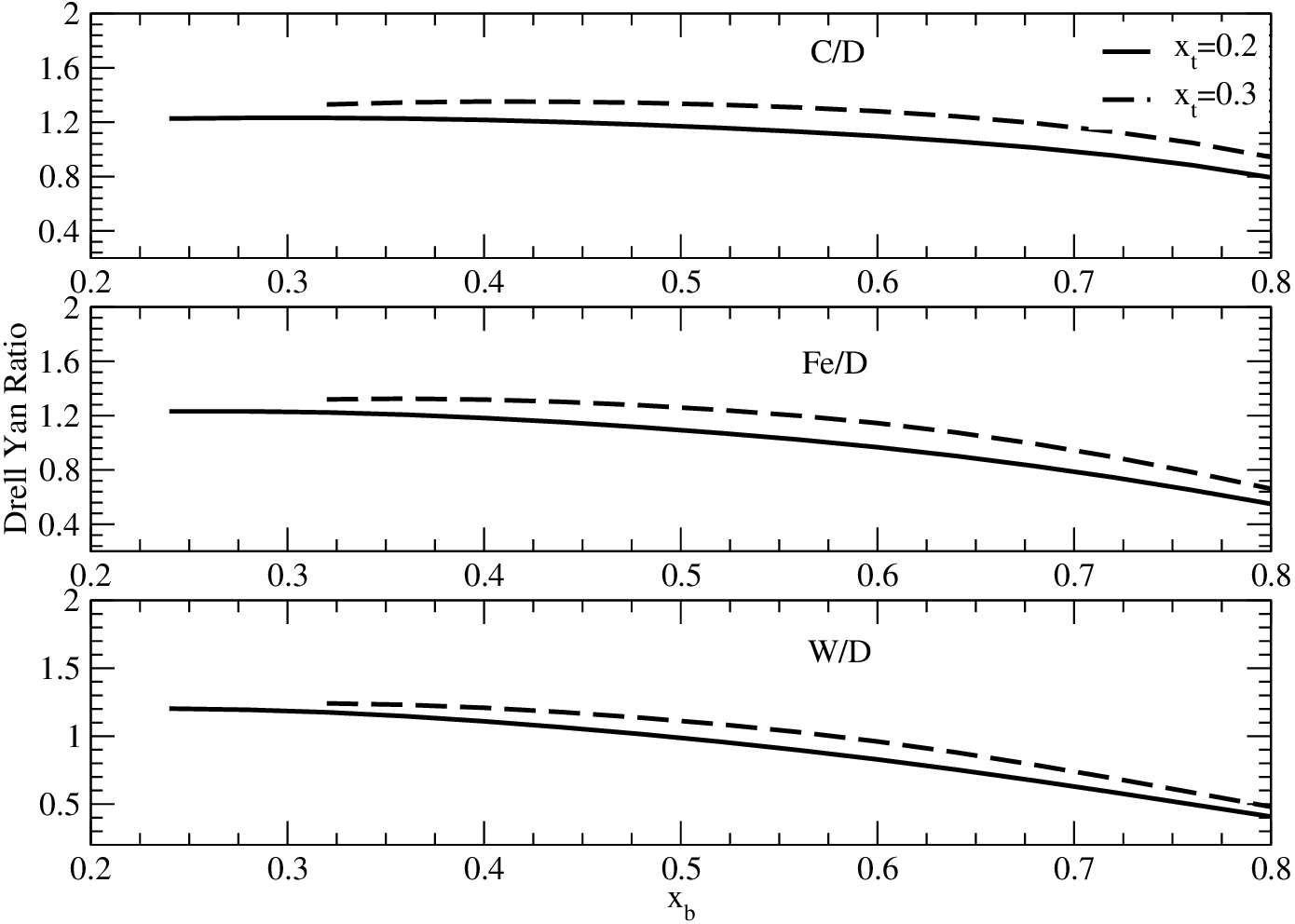}
\caption{$\frac{\left(\frac{d^2\sigma}{dx_bdx_t}\right)^{i}}{\left(\frac{d^2\sigma}{dx_bdx_t}\right)^{D}}$ vs $x_b$ 
at different $x_t$ for $E_{Lab}=120GeV$, $Q^2 > 5GeV^2$ and $\beta=0.04$. i stands for C, 
Fe and W nuclei.}\label{fig7}  
\end{figure}

Keeping the SeaQuest\cite{Seaquest} experiment at Fermi Lab in mind where various nuclear targets like deuterium, carbon, iron and tungsten are proposed to be used using a beam energy of 120GeV,
in Fig. \ref{fig7}, we present the results for the Drell-Yan ratio for $\frac{\left(\frac{d^2\sigma}{dx_bdx_t}\right)^{i}}{\left(\frac{d^2\sigma}{dx_bdx_t}\right)^{D}}$ vs $x_b$ for different values
 of $x_t$, for $Q^2 > 5GeV^2$ at $E_{Lab}=120GeV$, i stands for C, Fe and W nuclei.

\section{Summary and Conclusion}\label{sec:CC}
We have studied nuclear medium effects in Drell-Yan processes at small
target $x_t$ away from the shadowing region($x_t\ge0.1$) using quark parton distribution functions and nucleon structure functions for a bound nucleon. We have used a microscopic nuclear model which takes into account the effect of Fermi
motion, nuclear binding and nucleon correlations through a relativistic spectral function of bound nucleon. The contributions of $\pi$ and $\rho$ mesons are also included.
We also include the beam energy loss effect due to initial state interactions of protons visualized through inelastic collisions of protons with nuclear constituents before they suffer 
hard collisions to produce lepton pair. 
 We find a reduction in the DY yield due to nuclear structure effects and an enhancement due to mesonic contribution. Both the reduction as well as the enhancement in the case of DY yields are found to be 
 larger than found in the case of DIS of charged leptons for the same value of model parameters. In the case of DY yields there is a further reduction due to beam energy loss effect in the nuclear medium
 which has been treated using a parameter describing the beam energy loss. The numerical results are compared with 
 various theoretical results available in the literature and also with the experimental results from E772~\cite{alde} and E866~\cite{e866,vasilev} experiments.
 We find a reasonable agreement with the experimental results presently available for $^{12}C$, $^{40}Ca$, $^{56}Fe$, and $^{184}W$ by suitably varying the beam energy loss parameter. We have also presented in this paper, results for $\frac{d^{2}\sigma}{dx_bdx_t}$ vs $x_b$ for various
 values of $x_t$ and the results for  $\frac{d\sigma}{dx_t}$ vs $x_t$  relevant to the 
forthcoming E906 SeaQuest~\cite{Seaquest} experiment at Fermi Lab. Our results show that the model for describing the nuclear medium effects in the DIS of charged leptons and 
neutrino and antineutrino with nuclear targets is able to explain the experimental results in the case of Drell-Yan yield in the region $0.1 < x_t < 0.35$ provided a reasonable model 
for beam energy loss effect is used. High statistics, high precision data from E906 SeaQuest~\cite{Seaquest} 
experiment on $\frac{d^{2}\sigma}{dx_bdx_t}$ in various regions of $x_b$ and $x_t$ will provide 
important information about the modification of quark PDFs and nucleon structure function in the nuclear medium.
\section{Acknowledgements}
M. S. A. is thankful to Department of Science
and Technology(DST), Government of India for providing financial assistance under Grant No. SR/S2/HEP-18/2012. I. R. S. thanks FIS2011-24149 Spanish project for financial support.

\end{document}